\documentclass[12pt,a4paper,dvips]{article}
\usepackage{a4p}
\usepackage{cite,mcite}
\usepackage{graphicx}
\usepackage{physics }
\usepackage{l3_title,ifthen}
\usepackage{amsmath,rotating,color}
\journalname{Phys. Lett. B}
\date{December 17, 2002}
\preprint{2002-092}
\newlength{\capindent}
\newlength{\capwidth}
\newlength{\figwidth}
\newcommand{\icaption}[2][!*!,!]{\hspace*{\capindent}%
  \begin{minipage}{\capwidth}
    \ifthenelse{\equal{#1}{!*!,!}}%
      {\caption{#2}}%
      {\caption[#1]{#2}}
  \end{minipage}}
\newcommand{\pho}{\phantom{0}}

\newcommand{\enqq}{{\ensuremath{\text{e}\nu\text{q}\bar{\text{q}}'}}}

\newcommand{\mnqq}{{\ensuremath{\mu\nu\text{q}\bar{\text{q}}'}}}
\newcommand{\tnqq}{{\ensuremath{\tau\nu\text{q}\bar{\text{q}}'}}}

\newcommand{\tlep}{{\ensuremath{\theta^*_\ell}}}
\newcommand{\thad}{{\ensuremath{\theta^*_{\text{q}}}}}

\newcommand{\fm}{{\ensuremath{f_{-}}}}
\newcommand{\fp}{{\ensuremath{f_{+}}}}
\newcommand{\fpm}{{\ensuremath{f_{\pm}}}}
\newcommand{\fo}{{\ensuremath{f_{0}}}}

\newcommand{\costlep}{\ensuremath{\cos\tlep}}
\newcommand{\costhad}{\ensuremath{|\cos\thad|}}

\newcommand{\thetawm}{\ensuremath{\Theta_\Wm}}

\begin{document}
\begin{titlepage}
\title{Measurement of W Polarisation at LEP}
\author{The L3 Collaboration}
%
% The abstract
%
\begin{abstract}
The three different helicity states of W bosons produced in the reaction 
$\mathrm{e}^{+}\mathrm{e}^{-}~\rightarrow 
\mathrm{W}^{+}\mathrm{W}^{-}~\rightarrow~\mathrm{\ell\nu{}q\bar{q}'}$ at LEP
are studied using leptonic and hadronic W decays.
Data at centre-of-mass energies $\sqrt{s}$~=~$183-209$~\gev{} are used to measure
the polarisation of W bosons, and its dependence on the W boson production angle.
The fraction of longitudinally polarised W bosons is measured to be 
0.218~$\pm$~0.027~$\pm$~0.016 where the first uncertainty is statistical and 
the second systematic, in agreement with the Standard Model expectation.

\end{abstract}
%
% Adds "To be submitted to ..." or "Submitted to ...", if relevant
%
\submitted
\end{titlepage}
%
%%%%%%%%%%%%%%%%%%%%%%%%%%%%%%%%%%%%%%%%%%%%%%%%%%%%%%%%%%%%%%%%%%%%%%%%%%%%%%%
% Introduction
%%%%%%%%%%%%%%%%%%%%%%%%%%%%%%%%%%%%%%%%%%%%%%%%%%%%%%%%%%%%%%%%%%%%%%%%%%%%%%%
%
\section*{Introduction}
The existence of all three W boson helicity states, $+1$, $-1$ and $0$,
is a consequence of the non-vanishing mass of the W boson,
that, in the Standard Model~\cite{standard_model}, is generated by the Higgs
mechanism of electroweak symmetry breaking. 
The measurement of the fractions of longitudinally and transversely polarised
W bosons constitutes a test of the Standard Model predictions 
for the triple gauge boson couplings $\gamma$WW and ZWW.

To determine the W helicity fractions, events of the type \epem{}\ra{}\Wp\Wm{}\ra$\ell\nu\qqbar'$ 
are used, with $\ell$ denoting either an electron or a muon. 
These events are essentially background free and allow a measurement, 
with good accuracy, of the W momentum vector, 
the W charge and the polar decay angles.
The W helicity states are accessible in a model independent way through the 
shape of the distributions of the polar decay angle, \tlep{}, 
between the charged lepton and the W direction in the W rest frame.
Transversely polarised W bosons have angular distributions
$(1 \mp \cos \tlep)^{2}$ for a \Wm{} with helicity $\pm1$, 
and $(1 \pm \cos \tlep)^{2}$ for a \Wp{} with helicity $\pm1$. 
For longitudinally polarised W bosons, a $\sin^{2} \tlep$ dependence is expected.
For simplicity, we refer in the following only to the fractions 
\fm{}, \fp{} and \fo{} of the helicity states $-1$, $+1$ and 0 of the \Wm{} boson, respectively.
Assuming CP invariance these equal the fractions of the corresponding 
helicity states $+1$, $-1$ and 0 of the \Wp{} boson.

The differential distribution of leptonic \Wm{} decays at Born level is:
\begin{equation}
\frac{1}{N}\frac{dN}{d\cos\tlep{}} = \fm{} \frac{3}{8}~(1+\cos\tlep{})^{2} + 
\fp{} \frac{3}{8}~(1-\cos\tlep{})^{2} + \fo{} \frac{3}{4} \sin^{2} \tlep{}.
\end{equation}

For hadronic W decays, the quark charge is difficult to reconstruct experimentally and 
only the absolute value of the cosine of the decay angle, \costhad{}, is used:
\begin{equation}
\frac{1}{N}\frac{dN}{d\costhad} = \fpm{} \frac{3}{4} (1+\costhad{}^{2}) 
+ \fo{} \frac{3}{2} (1-\costhad{}^2),
\end{equation}
with \fpm{}=\fp{}+\fm{}.

After correcting the data for selection efficiencies and background, the different fractions of W helicity 
states are obtained from a fit to these distributions.
The fractions \fm{}, \fp{} and \fo{} are also determined 
as a function of the \Wm{} production angle \thetawm{} 
in the laboratory frame.
The helicity composition of the W bosons depends strongly on the centre-of-mass 
energy, \rts{}.

%
%%%%%%%%%%%%%%%%%%%%%%%%%%%%%%%%%%%%%%%%%%%%%%%%%%%%%%%%%%%%%%%%%%%%%%%%%%%%%%%
% Data and MC
%%%%%%%%%%%%%%%%%%%%%%%%%%%%%%%%%%%%%%%%%%%%%%%%%%%%%%%%%%%%%%%%%%%%%%%%%%%%%%%
%
\section*{Data and Monte Carlo}
The analysis presented in this Letter is based on the whole data set collected 
with the L3 detector~\cite{l3_00}
and supersedes our previous results \cite{wwlong} based on about one third of the data.
An integrated luminosity of 684.8~\pb{}, collected at different centre-of-mass energies
between 183~\gev{} and 209~\gev{}, as shown in Table~\ref{tab:table1}, is analysed.

The \epem{}\ra{}\Wp{}\Wm{}\ra{}\enqq{},~\mnqq{} Monte Carlo events are generated using
KORALW~\cite{KORALW1}.
The Standard Model predictions for \fm{}, \fp{} and \fo{} are obtained from these samples
by fitting the generated decay angular distributions for each value of \rts{}.
As an example, the expected fraction of longitudinally polarised W bosons changes from
0.271 at \rts{}~=~183~\gev{} to 0.223 at \rts{}~=~206~\gev{}.
The luminosity averaged Standard Model expectations for
\fm{}, \fp{} and \fo{} are 0.590, 0.169 and 0.241, respectively.

Background processes are generated using KORALW for W pair production decaying to other final states, and 
PYTHIA~\cite{PYTHIA} and KK2F~\cite{KK2F} for \epem\ra{}\qqbar$(\gamma)$.
For studies of systematic effects, signal events are also generated using 
EEWW~\cite{EEWW} and EXCALIBUR~\cite{excalibur}.
The L3 detector response is simulated with the GEANT~\cite{geant} and GEISHA~\cite{gheisha} packages.
Detector inefficiencies, as monitored during the data taking period, are included.

A large sample of signal events is generated using the EEWW Monte Carlo program. 
This program assigns, differently from KORALW, W helicities on an event-by-event basis 
but uses the zero-width approximation for the W boson and does not include higher order radiative corrections 
and interference terms.
The \Wm{} helicity fractions obtained from a fit to the generated decay angle distributions agree
with the input values.
A comparison of the fractions obtained from EEWW and YFSWW~\cite{YFSWW}, which 
includes improved $O(\alpha)$ corrections, with those obtained from KORALW also shows good agreement.
Therefore the Born level formul\ae{} (1) and (2) are applicable after radiative corrections. 

%
%%%%%%%%%%%%%%%%%%%%%%%%%%%%%%%%%%%%%%%%%%%%%%%%%%%%%%%%%%%%%%%%%%%%%%%%%%%%%%%
% Selection
%%%%%%%%%%%%%%%%%%%%%%%%%%%%%%%%%%%%%%%%%%%%%%%%%%%%%%%%%%%%%%%%%%%%%%%%%%%%%%%
%
\section*{Selection of W$^{+}$W$^-  \rightarrow$ e\boldmath{$\nu\text{q}\bar{\text{q}}', \mu\nu\text{q}\bar{\text{q}}'$}  events}

\begin{figure}[htbp]
  \begin{center}
    \includegraphics[width=8cm]{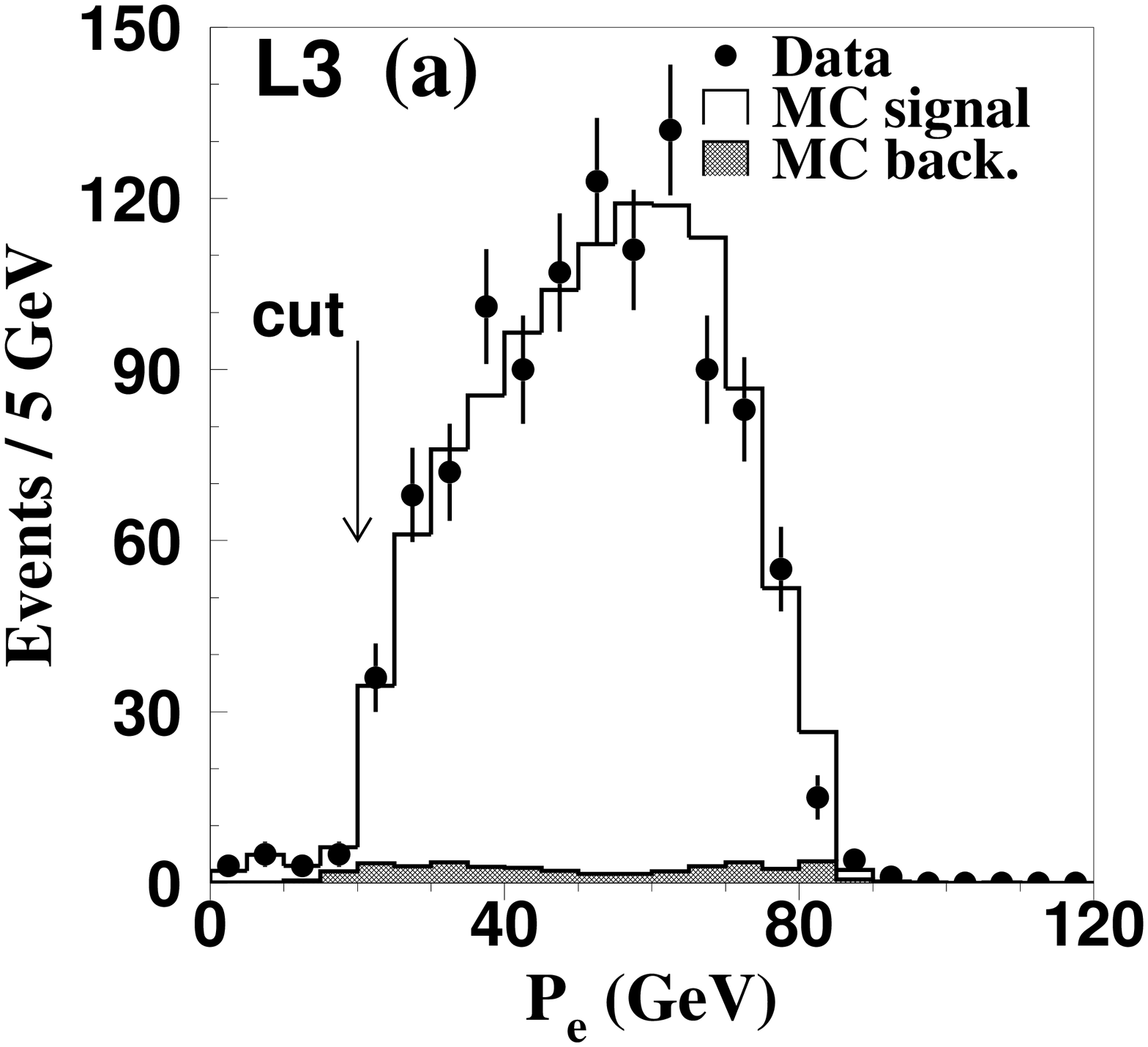}
    \includegraphics[width=8cm]{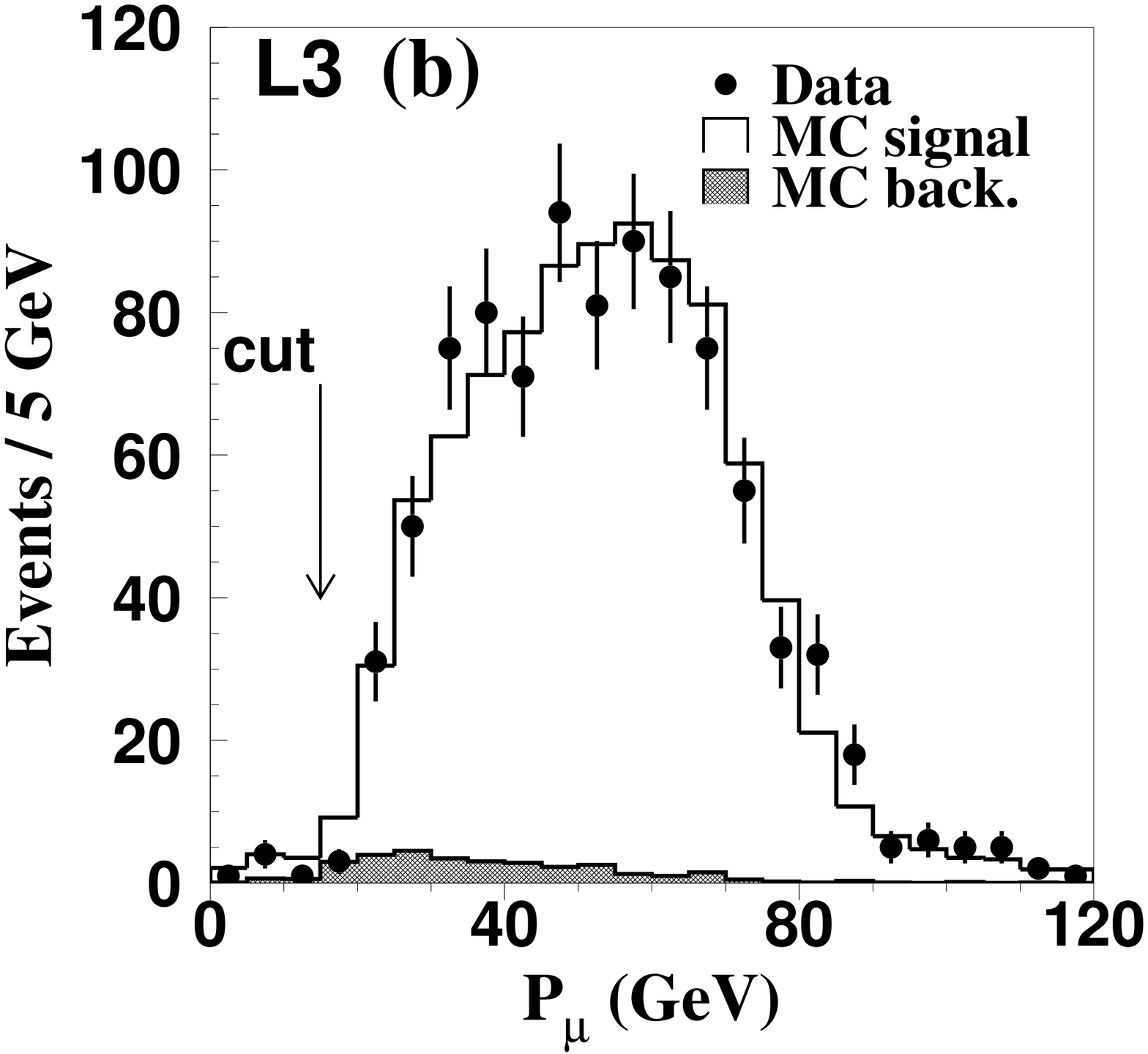}\\
    \includegraphics[width=8cm]{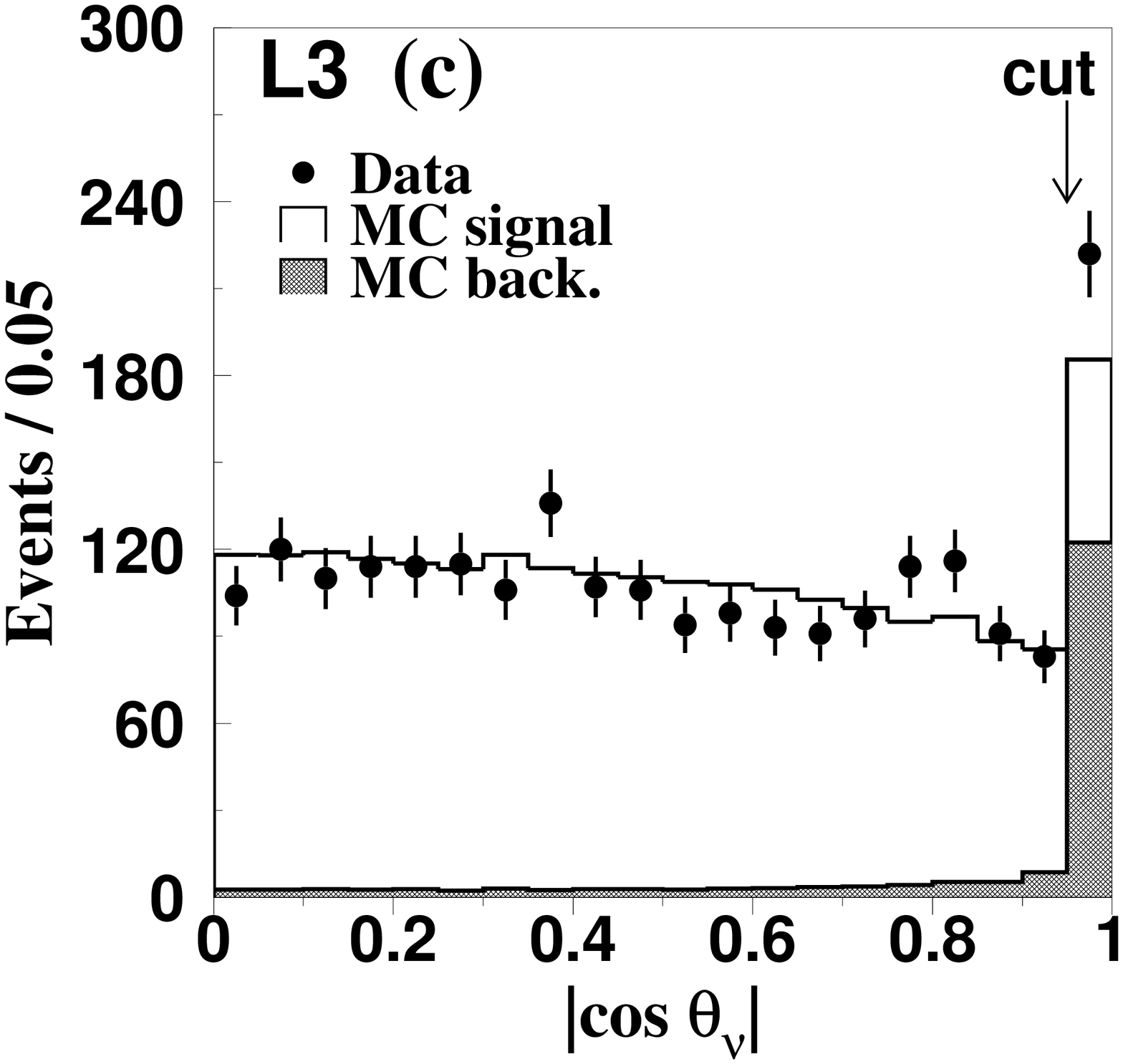}
    \includegraphics[width=8cm]{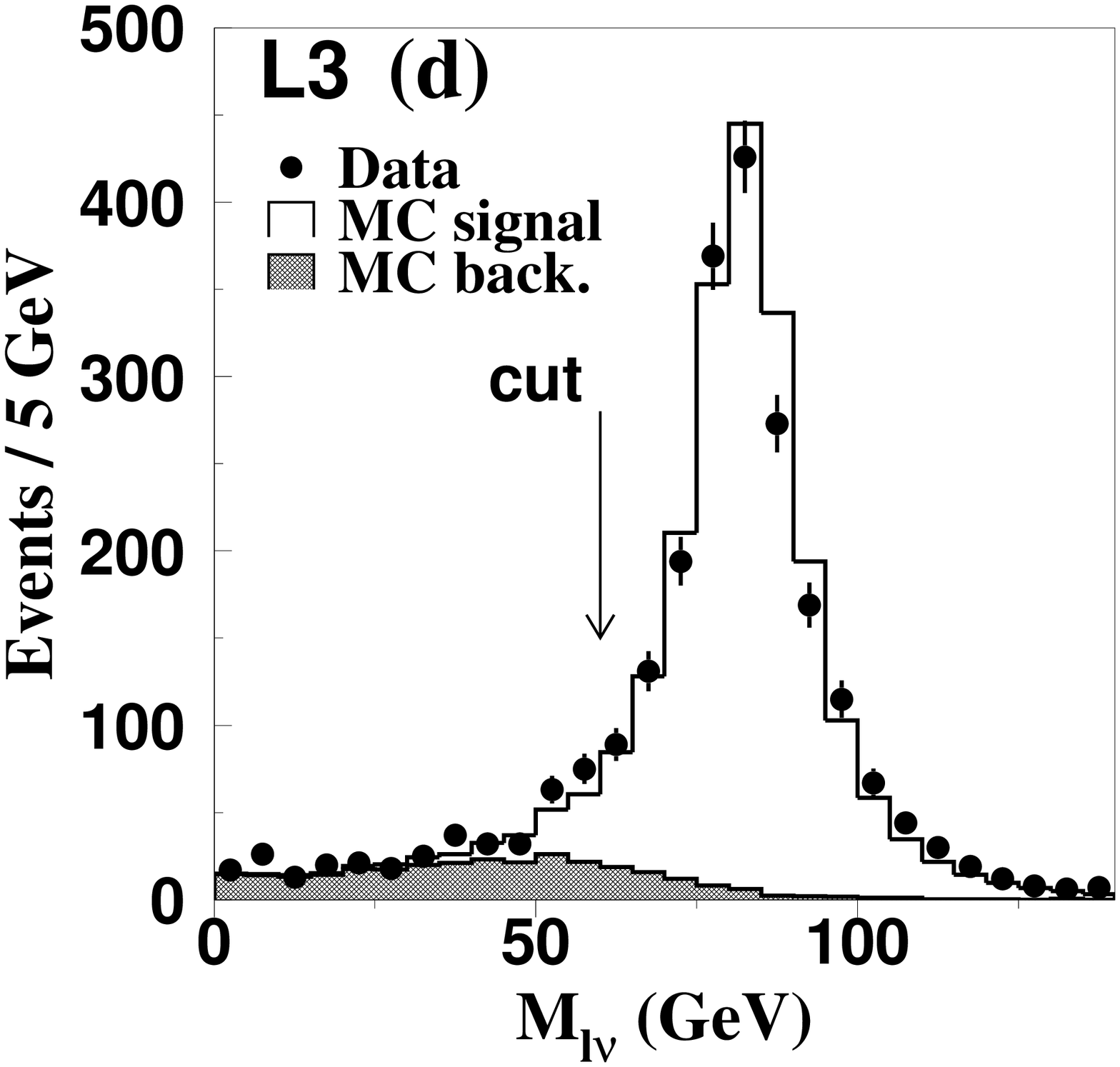}
  \end{center}
  \icaption{Distributions used for the event selections:
(a) momentum of the electrons,
(b) momentum of the muons,
(c) absolute value of the cosine of the polar angle of the neutrino,
(d) mass of the lepton-neutrino system.
In each plot, all other selection criteria are applied.
The arrows indicate cut positions.
\label{fig:figure4}}
\end{figure}

Only events which contain exactly one electron or one muon candidate are accepted~\cite{wwlong}.
Electrons are identified as isolated energy depositions in the electromagnetic calorimeter 
with an electromagnetic shower shape.
A match in azimuthal angle with a track reconstructed in the central tracking chamber is required.
Muons are identified and measured as tracks reconstructed in the muon chambers 
which point back to the interaction vertex. 
All other energy depositions in the calorimeters are assumed to 
originate from the hadronically decaying W. 
The neutrino momentum vector is assumed to be the missing momentum 
vector of the event. 
The following additional criteria are applied:
\begin{itemize}
\item The reconstructed momentum 
must be greater than 20 \GeV{} for electrons and 15 \GeV{} for muons. 
\item
The neutrino momentum must be greater than 10 \GeV{} 
and its polar angle, $\theta_{\nu}$, 
has to satisfy $|\cos \theta_{\nu}| < 0.95$.
\item 
The invariant mass of the lepton-neutrino system has to be 
greater than 60 \GeV{}.
\item
The invariant mass of the hadronic system has to be 
between 50 and 110 \GeV{}.
\end{itemize}
Figure \ref{fig:figure4} shows some distributions of those variables for data and Monte Carlo.

The number of events selected by these criteria are listed in 
Table~\ref{tab:table1}.
In total, 2010 events are selected with an efficiency of 65.7\% and a purity of 96.3\%.
The contamination from \Wp{}\Wm{}\ra{}\tnqq{} and \epem\ra{}\qqbar$(\gamma)$ is
2.4\% and 1.3\%, respectively, independent of \rts{} and the W
production angle.

%%%%%%%%%%%%%%%%%%%%%%%%%%%%%%%%%%%%%%%%%%%%%%%%%%%%%%%%%%%%%%%%%%%%%%%%%%%%%%%
% Analysis 
%%%%%%%%%%%%%%%%%%%%%%%%%%%%%%%%%%%%%%%%%%%%%%%%%%%%%%%%%%%%%%%%%%%%%%%%%%%%%%%
%
\section*{Analysis of the \bf{W} helicity states}
For the selected events, the rest frames of the W bosons are calculated 
from the lepton and neutrino momenta, and 
the decay angles \tlep{} and \thad{} of the lepton and the quarks are determined.
The angle \thad{} is approximated by the polar angle of the thrust axis 
with respect to the W direction in the rest frame 
of the hadronically decaying W.

The fractions of the W helicity states
are obtained from the event distributions, $dN/d\costlep{}$ and $dN/d\costhad{}$.
For each energy point, the background, as obtained from Monte Carlo simulations, 
is subtracted from the data, and the resulting distributions are corrected for 
selection efficiencies as obtained from large samples of KORALW Monte Carlo events. 
The corrected decay angle distributions at the different centre-of-mass 
energies are combined into single distributions for leptonic and hadronic decays, which are then
fitted to the functions (1) and (2), respectively. 
A binned fit is performed on the normalised distributions, 
shown in Figure~\ref{fig:figure1}, using \fm{} and \fo{} as the fit parameters.
The fraction \fp{} is obtained by constraining the sum of all three parameters to unity.

Detector resolution introduces migration effects that bias the fitted parameters.
For example, purely longitudinally polarised leptonically decaying W bosons 
at \rts{}~=~206~\gev{} would be measured to have a helicity composition:
\fo{}~=~0.945, \fm{}~=~0.043 and \fp{}~=~0.012.
The magnitude of these effects depends on the helicity fractions and on \rts{}.
Corrections for this bias as a function of the helicity fractions
are determined from EEWW Monte Carlo samples.
If the ratio of two helicity fractions is constant the bias correction function of 
the third fraction is linear to a good approximation.
For the correction of \fo{} in the hadronic W decay, the ratio \fm{}/\fp{} 
is taken from the measurement in the leptonic W decay, as only the 
sum of \fp{} and \fm{} is known from hadronic decays.
Bias correction functions are determined for the analysis of the complete data sample, 
separately for the \Wp{} and \Wm{} events and in bins of the \Wm{} production angle.

%%%%%%%%%%%%%%%%%%%%%%%%%%%%%%%%%%%%%%%%%%%%%%%%%%%%%%%%%%%%%%%%%%%%%%%%%%%%%%%
% Results and Systematics
%%%%%%%%%%%%%%%%%%%%%%%%%%%%%%%%%%%%%%%%%%%%%%%%%%%%%%%%%%%%%%%%%%%%%%%%%%%%%%%
%
\section*{Results}

\begin{figure}[htbp]
  \begin{center}
    \includegraphics[width=12cm]{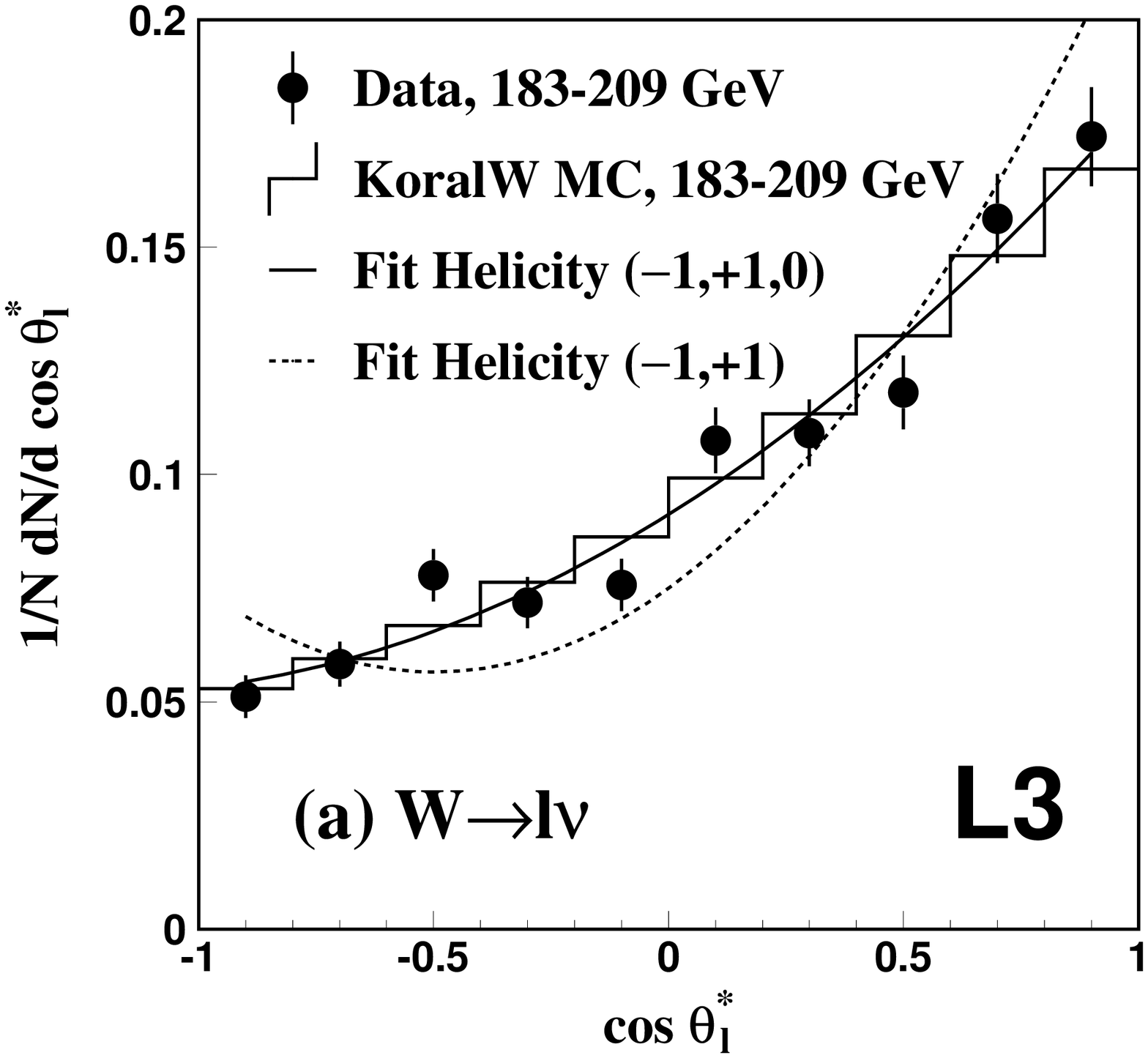}\\ \vspace{-15mm}
    \includegraphics[width=12cm]{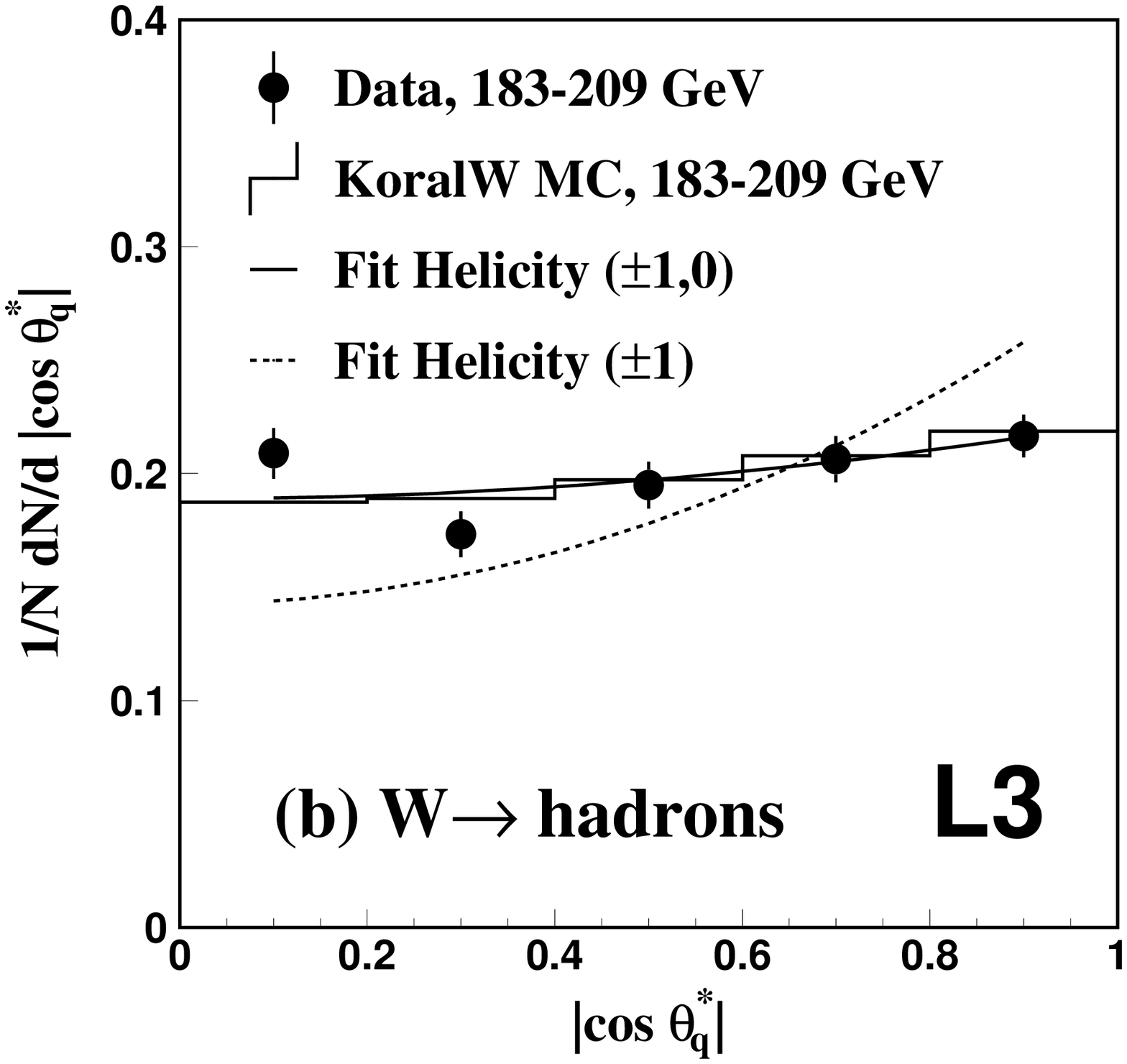}\\ \vspace{-10mm}
  \end{center}
  \icaption{Corrected decay angle distributions for (a) leptonic W decays
and (b) for hadronic W decays at \rts{}~=~$183-209$ \gev{}. 
Fit results for the different W helicity hypotheses are also shown.
\label{fig:figure1}}
\end{figure}

The results of the fits to the decay angle distributions for leptonic 
and hadronic W decays are shown in Figure~\ref{fig:figure1}.
The data are well described only if all three W helicity states are used.
Fits omitting the helicity 0 state fail to describe the data. 
For leptonic W decays, the $\chi^2$ increases 
from 12.7 for eight degrees of freedom if all helicity states are included 
to 56.2 for nine degrees of freedom if only the helicities $+1$ and $-1$ are used in the fit.
For hadronic W decays, the $\chi^2$ increases 
from 6.6 for four degrees of freedom if all helicity states are included 
to 59.1 for four degrees of freedom if only the helicities $\pm1$ are used.

The measured fractions of the W helicity states in data, at an average 
centre-of-mass energy \rts{}~=~196.7~\gev{}, are presented together with the Standard 
Model expectation in Tables~\ref{tab:table2a},~\ref{tab:table2b} and~\ref{tab:table3}.
The parameters \fm{} and \fo{} derived from the fit are about 90\% anti-correlated.
These results include a bias correction of 0.005 on \fo{} for leptonic decays 
and 0.044 for hadronic decays.
The bias correction adds 0.003 to the statistical uncertainty on \fo{}
for leptonic decays and 0.007 to the one for hadronic decays.
The measured W helicity fractions agree with the Standard Model expectations 
for the leptonic and hadronic decays, as well as for the combined sample.
Longitudinal W polarisation is observed with a significance of seven standard 
deviations, including systematic uncertainties.

A number of systematic uncertainties are considered.
These include selection criteria, binning effects, bias corrections, the contamination 
due to non double resonant four fermion processes, background levels, and efficiencies.
Selection cuts are varied over a range of one standard deviation of the 
corresponding reconstruction accuracy. 
Fits are repeated with one bin more or 
one bin less in the decay angle distributions.
Uncertainties on the bias and efficiency corrections are determined with large Monte Carlo
samples, the latter being negligible.
The contamination due to non double resonant four fermion processes is studied by 
using the EXCALIBUR Monte Carlo.
Background levels are varied according to Monte Carlo statistics for both the 
\epem\ra\Wp\Wm\ra\tnqq{} and \epem\ra\qqbar$(\gamma)$ processes.
The largest uncertainties arise from selection criteria and binning effects.
As an example, Table \ref{tab:table5} summarises those effects on \fo{}.

Within the Standard Model, CP symmetry is conserved in the reaction \epem\ra\Wp\Wm{} and 
the helicity fractions \fp{}, \fm{} and \fo{} for the \Wp{} are expected to be 
identical to the fractions \fm{}, \fp{} and \fo{}, for the \Wm{}, respectively.
CP invariance is tested by measuring the helicity fractions for 
\Wp{} and \Wm{} separately.
The charge of the W bosons is obtained from the charge of the lepton.
We select 1020 \Wp\ra$\ell^+\nu$, and 990 \Wm\ra$\ell^-\bar{\nu}$ events.
Results of separate fits for the \Wm{} helicity fractions are given in 
Tables~\ref{tab:table2a}, \ref{tab:table2b} and~\ref{tab:table3}
for leptonic, hadronic and combined fits.
Good agreement is found, consistent with CP invariance.

\begin{figure}[htbp]
  \begin{center}
    \includegraphics[width=\figwidth]{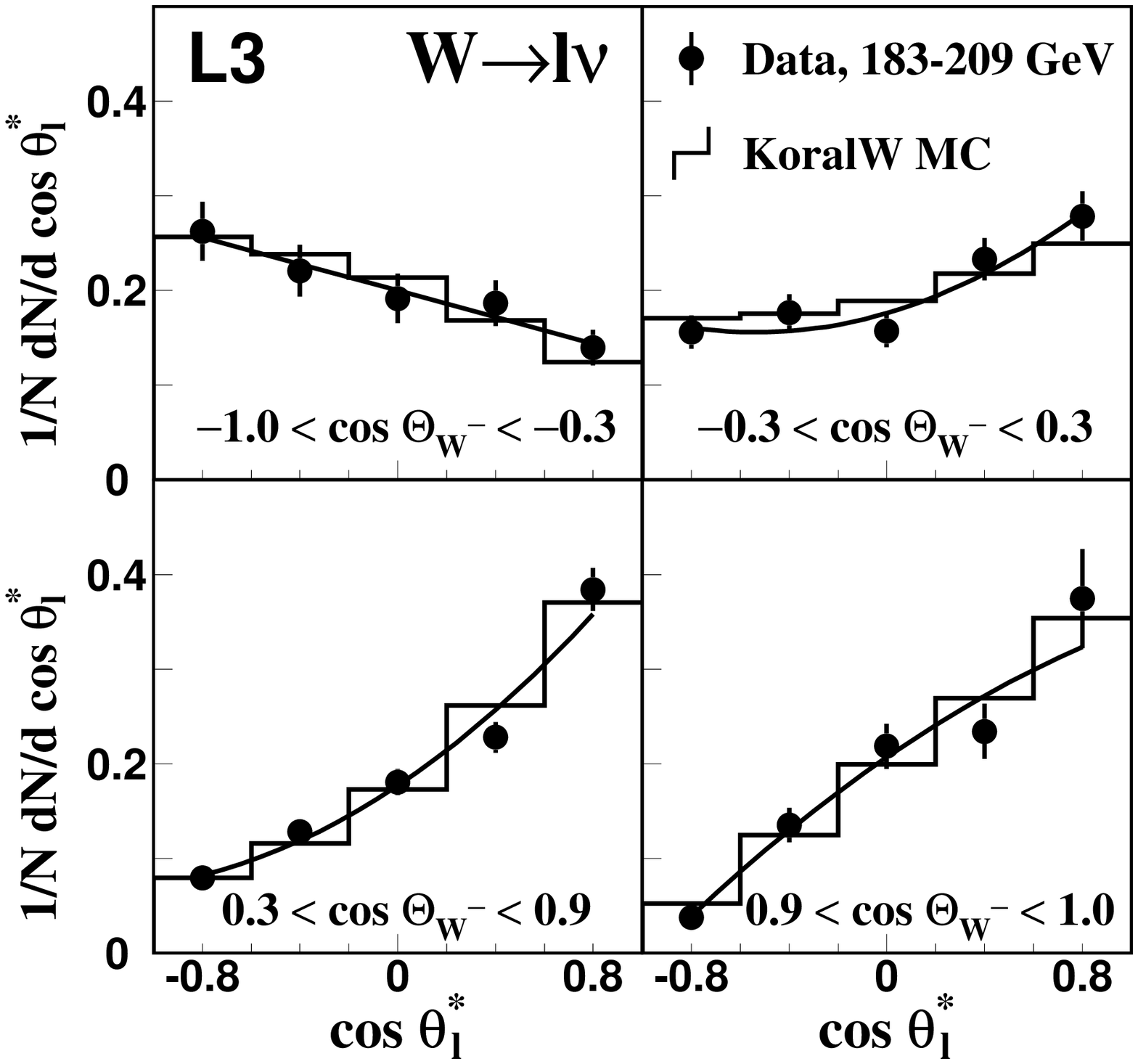}
  \end{center}
  \icaption{Corrected decay angle distributions for leptonic W decays separated into 
four different \Wm{} production angle ranges, together with the KORALW expectation
and fit results. Assuming CP invariance, \Wp{} decays are included with 
$\cos\thetawm{} = - \cos\Theta_{\Wp}$.
\label{fig:figure2}}
\end{figure}

\begin{figure}[htbp]
  \begin{center}
    \includegraphics[width=\figwidth]{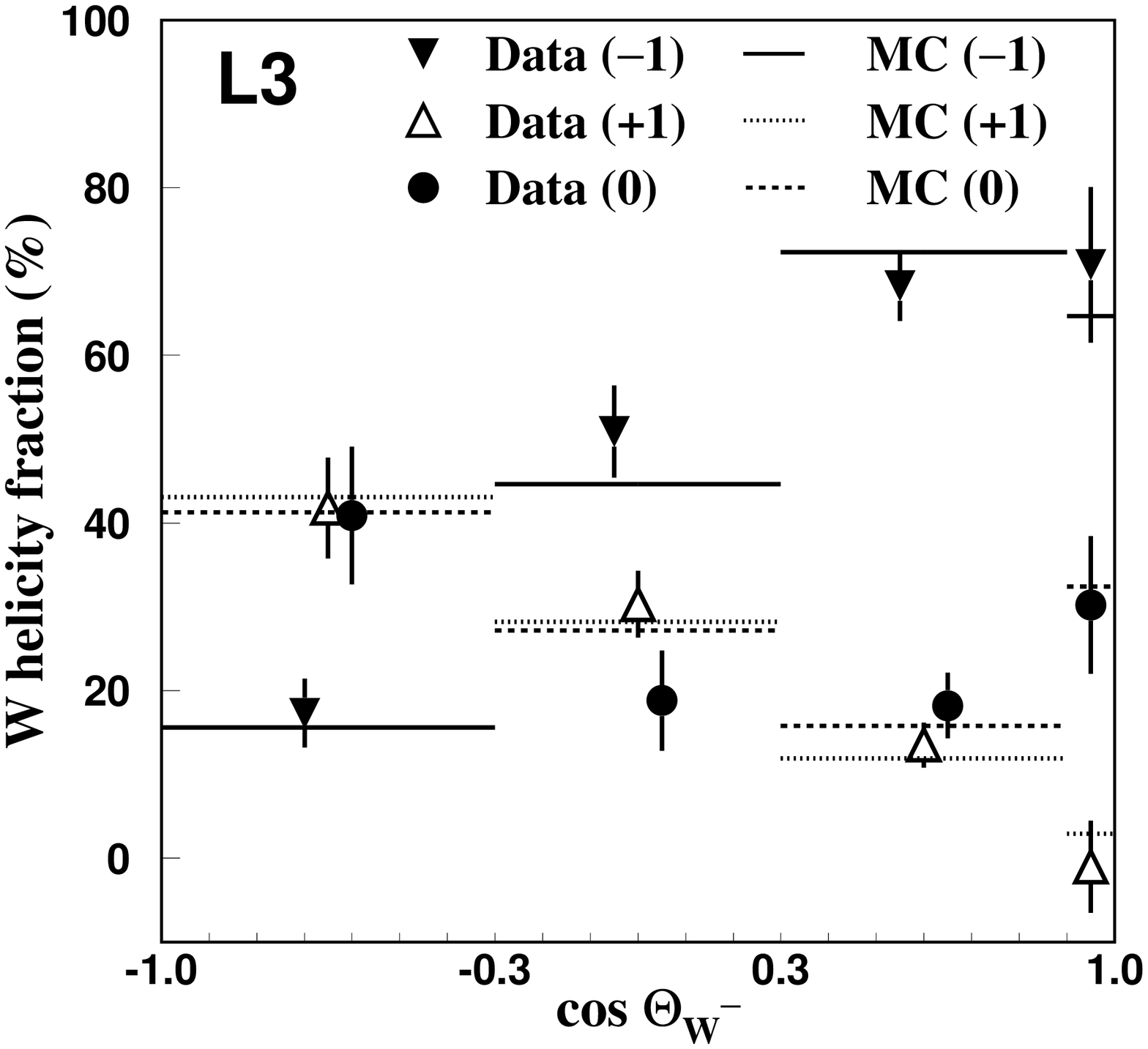}
  \end{center}
  \icaption{W helicity fractions $f_{-}$, $f_{+}$ and $f_{0}$ and their statistical uncertainties
for four different bins of $\cos \Theta_{\mathrm{W}^{-}}$
in the combined data sample and in the KORALW Monte Carlo for \rts{}~=~$183-209$ \gev{}.
\label{fig:figure3}}
\end{figure}

To test the variation of the helicity fractions with the \Wm{} 
production angle, \thetawm{}, the data are grouped in four bins of $\cos\thetawm{}$.
The ranges have been chosen such that large  and statistically significant variations of the different 
helicity fractions are expected.
Figure~\ref{fig:figure2} shows the four decay angle distributions for the leptonic W decays. 
The corrected distributions are fitted for leptonic and hadronic W decays separately 
in each bin of $\cos\thetawm{}$. 
The fit results, combining leptonic and hadronic W decays, 
are shown in Table~\ref{tab:table4} and Figure~\ref{fig:figure3}, together with the 
Standard Model expectations from the KORALW Monte Carlo. 
The results agree with the Standard Model expectation and demonstrate a strong variation 
of the W helicity fractions with the \Wm{} production angle.

In conclusion, all three helicity states of the W boson are required in order to describe the data.
Their fractions and their variations as a function of $\cos\thetawm{}$ are in agreement 
with the Standard Model expectation.
The fraction of longitudinally polarised W bosons at $\sqrt{s}$~=~$183-209$~\gev{} 
is measured as 0.218 $\pm$ 0.027 $\pm$ 0.016.
Separate analyses of the \Wp{} and \Wm{} events are consistent with CP conservation.

\bibliographystyle{l3stylem}

%
%%%%%%%%%%%%%%%%%%%%%%%%%%%%%%%%%%%%%%%%%%%%%%%%%%%%%%%%%%%%%%%%%%%%%%%%%%%%%%
% Author List
%%%%%%%%%%%%%%%%%%%%%%%%%%%%%%%%%%%%%%%%%%%%%%%%%%%%%%%%%%%%%%%%%%%%%%%%%%%%%%
%
\newpage
\typeout{   }     
\typeout{Using author list for paper 261 -  }
\typeout{$Modified: Jul 15 2001 by smele $}
\typeout{!!!!  This should only be used with document option a4p!!!!}
\typeout{   }
%
%
%
%  L A T E X  version!!
%
%
% Make sure that the Lep package has been used!
%\input{Lep.sty}%
%
%\ifx\LepCalled\undefined%
%\typeout{     }%
%\typeout{!!!!!!!!!!!!!!!!!!!!!!!!!!!!!!!!!!!!!!!!!!!!!!!!!!!!!!!!!!!}%
%\typeout{Yikes.  You haven't used the Lep package!}%
%\typeout{Please put \protect\usepackage\protect{Lep\protect} in your preamble,
%         followed by}%
%\typeout{\protect\Lep\protect{1\protect} or \protect\Lep\protect{2\protect}}%
%\typeout{     }%
%\typeout{For now you will get a Lep phase 2 authorlist (may not be right!).}%
%\typeout{!!!!!!!!!!!!!!!!!!!!!!!!!!!!!!!!!!!!!!!!!!!!!!!!!!!!!!!!!!!}%
%\typeout{     }%
%\Lep{2}\fi%

\newcount\tutecount  \tutecount=0
\def\tutenum#1{\global\advance\tutecount by 1 \xdef#1{\the\tutecount}}
\def\tute#1{$^{#1}$}
\tutenum\aachen            % 1
\tutenum\nikhef            % 2
\tutenum\mich              % 3
\tutenum\lapp              % 4
\tutenum\basel             % 5
\tutenum\lsu               % 6
\tutenum\beijing           % 7
\tutenum\bologna           % 8
\tutenum\tata              % 9 
\tutenum\ne                % 10
\tutenum\bucharest         % 11
\tutenum\budapest          % 12
\tutenum\mit               % 13
\tutenum\panjab            % 14 
\tutenum\debrecen          % 15
\tutenum\dublin            % 16
\tutenum\florence          % 17
\tutenum\cern              % 18
\tutenum\wl                % 19
\tutenum\geneva            % 20
\tutenum\hefei             % 21
\tutenum\lausanne          % 22
\tutenum\lyon              % 23
\tutenum\madrid            % 24
\tutenum\florida           % 25
\tutenum\milan             % 26
\tutenum\moscow            % 27
\tutenum\naples            % 29
\tutenum\cyprus            % 30
\tutenum\nymegen           % 31
\tutenum\caltech           % 32
\tutenum\perugia           % 33
\tutenum\peters            % 34
\tutenum\cmu               % 35
\tutenum\potenza           % 36
\tutenum\prince            % 37
\tutenum\riverside         % 38
\tutenum\rome              % 39
\tutenum\salerno           % 40
\tutenum\ucsd              % 41
\tutenum\sofia             % 42
\tutenum\korea             % 43
\tutenum\purdue            % 44
\tutenum\psinst            % 45
\tutenum\zeuthen           % 46
\tutenum\eth               % 47
\tutenum\hamburg           % 48
\tutenum\taiwan            % 49
\tutenum\tsinghua          % 50

{
\parskip=0pt
\noindent
{\bf The L3 Collaboration:}
\ifx\selectfont\undefined%  old style font selection
 \baselineskip=10.8pt
 \baselineskip\baselinestretch\baselineskip
 \normalbaselineskip\baselineskip
 \ixpt
\else%                      new style font selection
 \fontsize{9}{10.8pt}\selectfont
\fi
\medskip
\tolerance=10000
\hbadness=5000
\raggedright
\hsize=162truemm\hoffset=0mm
\def\r{\rlap,}
\noindent

P.Achard\r\tute\geneva\ 
O.Adriani\r\tute{\florence}\ 
M.Aguilar-Benitez\r\tute\madrid\ 
J.Alcaraz\r\tute{\madrid,\cern}\ 
G.Alemanni\r\tute\lausanne\
J.Allaby\r\tute\cern\
A.Aloisio\r\tute\naples\ 
M.G.Alviggi\r\tute\naples\
H.Anderhub\r\tute\eth\ 
V.P.Andreev\r\tute{\lsu,\peters}\
% ...
\newpage
%\rule{\textwidth}{0.4pt}
\begin{list}{A}{\itemsep=0pt plus 0pt minus 0pt\parsep=0pt plus 0pt minus 0pt
                \topsep=0pt plus 0pt minus 0pt}
\item[\aachen]
 III. Physikalisches Institut, RWTH, D-52056 Aachen, Germany$^{\S}$
\item[\nikhef] National Institute for High Energy Physics, NIKHEF, 
     and University of Amsterdam, NL-1009 DB Amsterdam, The Netherlands
\item[\mich] University of Michigan, Ann Arbor, MI 48109, USA
\item[\lapp] Laboratoire d'Annecy-le-Vieux de Physique des Particules, 
     LAPP,IN2P3-CNRS, BP 110, F-74941 Annecy-le-Vieux CEDEX, France
\item[\basel] Institute of Physics, University of Basel, CH-4056 Basel,
     Switzerland
\item[\lsu] Louisiana State University, Baton Rouge, LA 70803, USA
\item[\beijing] Institute of High Energy Physics, IHEP, 
  100039 Beijing, China$^{\triangle}$ 
\item[\bologna] University of Bologna and INFN-Sezione di Bologna, 
     I-40126 Bologna, Italy
\item[\tata] Tata Institute of Fundamental Research, Mumbai (Bombay) 400 005, India
\item[\ne] Northeastern University, Boston, MA 02115, USA
\item[\bucharest] Institute of Atomic Physics and University of Bucharest,
     R-76900 Bucharest, Romania
\item[\budapest] Central Research Institute for Physics of the 
     Hungarian Academy of Sciences, H-1525 Budapest 114, Hungary$^{\ddag}$
\item[\mit] Massachusetts Institute of Technology, Cambridge, MA 02139, USA
\item[\panjab] Panjab University, Chandigarh 160 014, India.
\item[\debrecen] KLTE-ATOMKI, H-4010 Debrecen, Hungary$^\P$
\item[\dublin] Department of Experimental Physics,
  University College Dublin, Belfield, Dublin 4, Ireland
\item[\florence] INFN Sezione di Firenze and University of Florence, 
     I-50125 Florence, Italy
\item[\cern] European Laboratory for Particle Physics, CERN, 
     CH-1211 Geneva 23, Switzerland
\item[\wl] World Laboratory, FBLJA  Project, CH-1211 Geneva 23, Switzerland
\item[\geneva] University of Geneva, CH-1211 Geneva 4, Switzerland
\item[\hefei] Chinese University of Science and Technology, USTC,
      Hefei, Anhui 230 029, China$^{\triangle}$
\item[\lausanne] University of Lausanne, CH-1015 Lausanne, Switzerland
\item[\lyon] Institut de Physique Nucl\'eaire de Lyon, 
     IN2P3-CNRS,Universit\'e Claude Bernard, 
     F-69622 Villeurbanne, France
\item[\madrid] Centro de Investigaciones Energ{\'e}ticas, 
     Medioambientales y Tecnol\'ogicas, CIEMAT, E-28040 Madrid,
     Spain${\flat}$ 
\item[\florida] Florida Institute of Technology, Melbourne, FL 32901, USA
\item[\milan] INFN-Sezione di Milano, I-20133 Milan, Italy
\item[\moscow] Institute of Theoretical and Experimental Physics, ITEP, 
     Moscow, Russia
\item[\naples] INFN-Sezione di Napoli and University of Naples, 
     I-80125 Naples, Italy
\item[\cyprus] Department of Physics, University of Cyprus,
     Nicosia, Cyprus
\item[\nymegen] University of Nijmegen and NIKHEF, 
     NL-6525 ED Nijmegen, The Netherlands
\item[\caltech] California Institute of Technology, Pasadena, CA 91125, USA
\item[\perugia] INFN-Sezione di Perugia and Universit\`a Degli 
     Studi di Perugia, I-06100 Perugia, Italy   
\item[\peters] Nuclear Physics Institute, St. Petersburg, Russia
\item[\cmu] Carnegie Mellon University, Pittsburgh, PA 15213, USA
\item[\potenza] INFN-Sezione di Napoli and University of Potenza, 
     I-85100 Potenza, Italy
\item[\prince] Princeton University, Princeton, NJ 08544, USA
\item[\riverside] University of Californa, Riverside, CA 92521, USA
\item[\rome] INFN-Sezione di Roma and University of Rome, ``La Sapienza",
     I-00185 Rome, Italy
\item[\salerno] University and INFN, Salerno, I-84100 Salerno, Italy
\item[\ucsd] University of California, San Diego, CA 92093, USA
\item[\sofia] Bulgarian Academy of Sciences, Central Lab.~of 
     Mechatronics and Instrumentation, BU-1113 Sofia, Bulgaria
\item[\korea]  The Center for High Energy Physics, 
     Kyungpook National University, 702-701 Taegu, Republic of Korea
\item[\purdue] Purdue University, West Lafayette, IN 47907, USA
\item[\psinst] Paul Scherrer Institut, PSI, CH-5232 Villigen, Switzerland
\item[\zeuthen] DESY, D-15738 Zeuthen, Germany
\item[\eth] Eidgen\"ossische Technische Hochschule, ETH Z\"urich,
     CH-8093 Z\"urich, Switzerland
\item[\hamburg] University of Hamburg, D-22761 Hamburg, Germany
\item[\taiwan] National Central University, Chung-Li, Taiwan, China
\item[\tsinghua] Department of Physics, National Tsing Hua University,
      Taiwan, China
\item[\S]  Supported by the German Bundesministerium 
        f\"ur Bildung, Wissenschaft, Forschung und Technologie
\item[\ddag] Supported by the Hungarian OTKA fund under contract
numbers T019181, F023259 and T037350.
\item[\P] Also supported by the Hungarian OTKA fund under contract
  number T026178.
\item[$\flat$] Supported also by the Comisi\'on Interministerial de Ciencia y 
        Tecnolog{\'\i}a.
\item[$\sharp$] Also supported by CONICET and Universidad Nacional de La Plata,
        CC 67, 1900 La Plata, Argentina.
\item[$\triangle$] Supported by the National Natural Science
  Foundation of China.
\end{list}
}
\vfill

%%% Local Variables: 
%%% mode: latex
%%% TeX-master: t
%%% End:

\newpage

%%%%%%%%%%%%%%%%%%%%%%%%%%%%%%%%%%%%%%%%%%%%%%%%%%%%%%%%%%%%%%%%%%%%%%%%%%%%%%
% TABLES
%%%%%%%%%%%%%%%%%%%%%%%%%%%%%%%%%%%%%%%%%%%%%%%%%%%%%%%%%%%%%%%%%%%%%%%%%%%%%%
\begin{table}[htbp]
 \begin{center}
  \begin{tabular}{|c|c|c|c|c|c|c|c|}
    \hline
    $\langle\rts{}\rangle$ [\gev{}]& 182.7 & 188.6 & 191.6 & 195.5 & 199.5 & 201.8 & 205.9 \\
    \hline
    Integrated luminosity [\pb{}] & \pho{}55.5 & 176.8 & \pho{}29.8 &
        \pho{}84.1 & \pho{}83.3 & \pho{}37.2 & 218.1 \\
    \hline
    Selected \enqq{} events & \pho{}82 & 293 & \pho{}59 &  133 &  110 &
        \pho{}56 & 355 \\
    Selected \mnqq{} events & \pho{}67 & 255 & \pho{}43 & 110 & \pho{}99 &
        \pho{}59 & 289  \\
    \hline 
    \end{tabular}
    \icaption{Average centre-of-mass energies, integrated luminosities and numbers of selected events.
    \label{tab:table1}}
  \end{center}
\end{table}

\begin{table}[htbp]
  \begin{center}
    \begin{tabular}{|c|c|c|c|}\hline
         Sample & \fm{} & \fp{} & \fo{} \\
       \hline
        \Wm{}\ra{}$\ell^{-}\nu$ Data  & 0.559 $\pm$ 0.038 $\pm$ 0.016 & 0.201 $\pm$ 0.026 $\pm$ 0.015 & 0.240 $\pm$ 0.051 $\pm$ 0.017 \\
        \Wp{}\ra{}$\ell^{+}\nu$ Data  & 0.625 $\pm$ 0.037 $\pm$ 0.016 & 0.179 $\pm$ 0.023 $\pm$ 0.015 & 0.196 $\pm$ 0.050 $\pm$ 0.017 \\
\hline
        \Wpm{}\ra{}$\ell^\pm\nu$ Data & 0.589 $\pm$ 0.027 $\pm$ 0.016 & 0.189 $\pm$ 0.017 $\pm$ 0.015 & 0.221 $\pm$ 0.036 $\pm$ 0.017 \\
        Monte Carlo    & 0.592 $\pm$ 0.003 & 0.170 $\pm$ 0.002 & 0.238 $\pm$ 0.004 \\  \hline
    \end{tabular}
    \icaption{\Wm{} helicity fractions for the leptonic decays for the combined data sample. 
All the helicities are converted to \Wm{} parameters using CP invariance.
The first uncertainty is statistical, the second systematic.
The corresponding helicity fractions in the Standard Model 
as implemented in the KORALW Monte Carlo program 
are also given with their statistical uncertainties.
    \label{tab:table2a}}
  \end{center}
\end{table}

\begin{table}[htbp]
  \begin{center}
    \begin{tabular}{|c|c|c|}
        \hline
        Sample &  \fpm{} & \fo{} \\
       \hline
        \Wm{}\ra{}hadrons Data   & 0.750 $\pm$ 0.056 $\pm$ 0.039 & 0.250 $\pm$ 0.056 $\pm$ 0.039 \\
        \Wp{}\ra{}hadrons Data   & 0.833 $\pm$ 0.062 $\pm$ 0.039 & 0.167 $\pm$ 0.062 $\pm$ 0.039 \\
\hline
        \Wpm{}\ra{}hadrons Data  & 0.785 $\pm$ 0.042 $\pm$ 0.039 & 0.215 $\pm$ 0.042 $\pm$ 0.039 \\
        Monte Carlo           & 0.757 $\pm$ 0.004 & 0.243 $\pm$ 0.004 \\ \hline
    \end{tabular}
    \icaption{\Wm{} helicity fractions for the hadronic decays for the combined data sample. 
All the helicities are converted to \Wm{} parameters using CP invariance.
The first uncertainty is statistical, the second systematic.
The corresponding helicity fractions in the Standard Model 
as implemented in the KORALW Monte Carlo program 
are also given with their statistical uncertainties.
    \label{tab:table2b}}
  \end{center}
\end{table}

\begin{table}[htbp]
  \begin{center}
    \begin{tabular}{|c|c|c|c|}\hline
         & \fm{} & \fp{} & \fo{} \\
      \hline
        \Wm{} Data  & 0.555 $\pm$ 0.037 $\pm$ 0.016 & 0.200 $\pm$ 0.026 $\pm$ 0.015 & 0.245 $\pm$ 0.038 $\pm$ 0.016 \\
        \Wp{} Data  & 0.634 $\pm$ 0.038 $\pm$ 0.016 & 0.181 $\pm$ 0.024 $\pm$ 0.015 & 0.185 $\pm$ 0.039 $\pm$ 0.016 \\
\hline
        \Wpm{} Data & 0.592 $\pm$ 0.027 $\pm$ 0.016 & 0.190 $\pm$ 0.017 $\pm$ 0.015 & 0.218 $\pm$ 0.027 $\pm$ 0.016 \\
\hline
        Monte Carlo    & 0.590 $\pm$ 0.003 & 0.169 $\pm$ 0.002 & 0.241 $\pm$ 0.003 \\  \hline
    \end{tabular}
    \icaption{\Wm{} helicity fractions, measured combining leptonic and hadronic decays.
All the helicities are converted to \Wm{} parameters using CP invariance.
The first uncertainty is statistical, the second systematic.
The corresponding helicity fractions in the Standard Model 
as implemented in the KORALW Monte Carlo program 
are also given with their statistical uncertainties.
    \label{tab:table3}}
  \end{center}
\end{table}

\begin{table}[htbp]
  \begin{center}
    \begin{tabular}{|l|c|c|}\hline
          & W\ra{}$\ell\nu$ & W\ra{}hadrons \\
        \hline
        Selection                       & 0.013 & 0.024 \\
        Binning effects                 & 0.007 & 0.029 \\
        Bias correction                 & 0.006 & 0.011 \\
        Four fermion contamination      & 0.005 & 0.001 \\
        Background corrections          & 0.004 & 0.001 \\
        \hline
        Total & 0.017 & 0.039  \\
        \hline
    \end{tabular}
    \icaption{Systematic uncertainties on the measurement of \fo{} for leptonic and hadronic W decays.
    \label{tab:table5}}
  \end{center}
\end{table}

\begin{table}[htbp]
  \begin{center}
    \begin{tabular}{|c|c|c|c|}\hline
     $\cos \Theta_{\mathrm{W}^{-}}$ & Fraction & \Wpm{} Data & Monte Carlo \\
    \hline
                                       & \fm{} & \pho{} 0.173 $\pm$ 0.041 $\pm$ 0.033 & 0.156 $\pm$ 0.006 \\
    $[-1.0, -0.3] $                     & \fp{} & \pho{} 0.418 $\pm$ 0.060 $\pm$ 0.043 & 0.431 $\pm$ 0.008 \\
                                       & \fo{} & \pho{} 0.409 $\pm$ 0.082 $\pm$ 0.051 & 0.413 $\pm$ 0.008 \\
    \hline
                                       & \fm{} & \pho{} 0.509 $\pm$ 0.055 $\pm$ 0.029 & 0.446 $\pm$ 0.006 \\
    $[-0.3, \phantom{-}0.3]$            & \fp{} & \pho{} 0.303 $\pm$ 0.040 $\pm$ 0.032 & 0.282 $\pm$ 0.005 \\
                                       & \fo{} & \pho{} 0.188 $\pm$ 0.060 $\pm$ 0.043 & 0.272 $\pm$ 0.006 \\
    \hline
                                       & \fm{} & \pho{} 0.683 $\pm$ 0.042 $\pm$ 0.026 & 0.723 $\pm$ 0.004 \\
    $[\phantom{-}0.3, \phantom{-}0.9]$ & \fp{} & \pho{} 0.135 $\pm$ 0.027 $\pm$ 0.030 & 0.119 $\pm$ 0.003 \\
                                       & \fo{} & \pho{} 0.182 $\pm$ 0.039 $\pm$ 0.027 & 0.158 $\pm$ 0.004 \\
    \hline
                                       & \fm{} & \pho{} 0.708 $\pm$ 0.093 $\pm$ 0.056 & 0.647 $\pm$ 0.007 \\
    $[\phantom{-}0.9, \phantom{-}1.0]$ & \fp{} & $-$0.010 $\pm$ 0.055 $\pm$ 0.028 & 0.029 $\pm$ 0.004 \\
                                       & \fo{} & \pho{} 0.302 $\pm$ 0.082 $\pm$ 0.059 & 0.324 $\pm$ 0.007 \\
    \hline
    \end{tabular}
    \icaption{The \Wm{} helicity fractions measured as a function of 
$\cos \Theta_{\mathrm{W}^{-}}$ combining leptonic and hadronic W decays.
The first uncertainty is statistical, the second systematic.
The KORALW Monte Carlo expectations are also given with their statistical uncertainties.
    \label{tab:table4}}
  \end{center}
\end{table}

%%%%%%%%%%%%%%%%%%%%%%%%%%%%%%%%%%%%%%%%%%%%%%%%%%%%%%%%%%%%%%%%%%%%%%%%%%%%%%
% FIGURES
%%%%%%%%%%%%%%%%%%%%%%%%%%%%%%%%%%%%%%%%%%%%%%%%%%%%%%%%%%%%%%%%%%%%%%%%%%%%%%
\clearpage
\newpage

\newpage

\newpage

\newpage

\end{document}